# Pressure-Induced Structural Phase Transition and a Special Amorphization Phase of Two-Dimensional Ferromagnetic Semiconductor Cr$_2$Ge$_2$Te$_6$

Zhenhai Yu[†,#], Wei Xia[†,‡,#], Kailang Xu[§,#], Ming Xu[§,*], Hongyuan Wang[†,‡], Xia Wang[∥], Na Yu[∥], Zhiqiang Zou[∥], Jinggeng Zhao[⊥,*], Lin Wang[⊥], Xiangshui Miao[§], and Yanfeng Guo[†,‡,*]

[†]School of Physical Science and Technology, ShanghaiTech University, Shanghai 201210, China

[‡]University of Chinese Academy of Sciences, Beijing 100049, China

[§]School of Optical and Electronic Information, Huazhong University of Science and Technology, Wuhan 430074, China

[∥]Analytical Instrumentation Center, School of Physical Science and Technology, ShanghaiTech University, Shanghai 201210, China

[⊥]Department of Physics, Harbin Institute of Technology, Harbin 150080, China

[⊥]Center for High Pressure Science and Technology Advanced Research, Shanghai, 201203, China

**ABSTRACT:** Layered transition-metal trichalcogenides have become one of the research frontiers as two-dimensional magnets and candidate materials used for phase-change memory devices. Herein we report the high-pressure synchrotron X-ray diffraction and resistivity measurements on Cr$_2$Ge$_2$Te$_6$ (CGT) single crystal by using diamond anvil cell techniques, which reveal a mixture of crystalline-to-crystalline and crystalline-to-amorphous transitions taking place concurrently at 18.3-29.2 GPa. The polymorphic transition could be interpreted by atomic layer reconstruction and the amorphization could be understood in connection with randomly flipping atoms into van der Waals gaps. The amorphous (AM) phase is quenchable to ambient conditions. The electrical resistance of CGT shows a bouncing point at ~ 18 GPa, consistent with the polymorphism phase transition. Interestingly, the high-pressure AM phase exhibits metallic resistance with the magnitude comparable to that of high-pressure crystalline phases, whereas the resistance of the AM phase at ambient pressure fails to exceed that of the crystalline phase, indicating that the AM phase of CGT appeared under high pressure is quite unique and similar behavior has never been observed in other phase-change materials. The results definitely would have significant implications for the design of new functional materials.

## Introduction

Since the successful exfoliation of single-layer grapheme which displays intriguing physical properties that could be widely used in technology, two-dimensional (2D) materials have been attracting significant interest.[1] However, the absence of magnetism forbids them from the application in spintronic devices, which require researchers to seek alternative 2D materials.[2] Recently, 2D-like ferromagnetic materials such as Cr$_2$Ge$_2$Te$_6$ (CGT) have been successfully exfoliated even into a monolayer which surprisingly retains long-range ferromagnetic order at finite temperature.[3] This discovery has sparked tremendous interest in chromium tellurides Cr$_2$X$_2$Te$_6$ ($X$ = Si, Ge, and Sn) with the centrosymmetric structure as they belong to a rare category of ferromagnetic semiconductors possessing a 2D layered structure as shown in Figure 1 (a).[4,5] The emergence of ferromagnetism in 2D materials combined with their rich electronic and optical properties opens up numerous opportunities for 2D magneto-electric devices and magneto-optic applications. For example, Ji et al. reported the growth of topological insulators Bi$_2$Te$_3$ on a ferromagnetic insulating CGT substrate via metal-organic chemical vapor deposition (MOCVD),[6] which offers a possibility of studying the anomalous quantum Hall effect in topological insulators.

CGT is also a class of phase-change memory materials (PCMs) enabled by a large resistance contrast between amorphous and crystalline phases upon reversible switching between them.[7-12] Unlike most PCMs in which the AM phase has larger resistivity than the crystal, Hatayama et al. observed an inverse resistance change in CGT that shows a high-resistance crystalline reset state and a low-resistance amorphous set state by annealing the as-deposited films at elevated temperatures.[13] Recently, the influence of the N-dopant on the phase stability, electrical behavior, and device performance of CGT was investigated experimentally by Shuang et al.[14] The experimental results show that the crystalline phase has larger resistance by one order-of-magnitude after crystallization than the AM state.

Pressure as a conventional thermodynamic parameter is a clean and useful tool to tune the atomic/molecular distance and consequently is capable of affecting the physical properties. In different materials, pressure induces electronic and structural transformations such as crystalline-to-crystalline (polymorphism), amorphous-to-crystal (crystallization), and crystalline-to-amorphous (amorphization) transitions. The pressure-induced polymorphic transition has been successfully studied by synchrotron X-ray diffraction (XRD) for many chalcogenides such as topological insulator $Sb_2Te_3$,[15] iron-based superconductor FeSe,[16] and thermoelectric material PbSe.[17] However, the pressure-induced amorphization (PIA) was much less observed because it involves a large number of disordered defects and distortions. The effects of hydrostatic pressure on spin-lattice coupling in 2D ferromagnetic CGT were investigated by Sun *et al*. using both experiments (Raman spectroscopy and magnetic transport) and first principles calculations.[18] The pressure-induced increase of Raman modes without structure transition was observed (up to 5.71 GPa). The magnetic transport measurement showed that the magnetic phase transition temperature $T_c$ decreases from 66.6 to 60.6 K as the pressure increases from 0 to 1 GPa. Since the pressure plays such an effective role in influencing the magnetic interactions, it definitely would be valuable to investigate CGT to a much higher pressure to see the effects on the structural and electronic transport properties.

**Experimental Section**
**Crystal Growth**

The CGT crystals were grown by using a self-flux similar to that reported earlier.[4] Platelike black crystals with shining surface and a typical size of $5\times3\times0.2$ $mm^3$ were finally left, shown by the optical microscopy picture in Figure1(b).

**Crystal Characterization**

The crystallographic phase quality of the crystals was examined on a Bruker D8 single crystal X-ray diffractometer with $\lambda$ = 0.71073 Å at room temperature. The XRD measurements revealed the rhombohedral structure of all randomly selected crystals.

**High Pressure Synchrotron Angle Dispersive XRD (AD-XRD)**

Selected high-quality crystals were ground in a mortar in order to obtain a fine powder sample used for the following high pressure XRD and resistance measurements. The high-pressure synchrotron XRD experiments were performed using a symmetric diamond anvil cell (DAC) with 300 μm culet diamond. The sample chambers were filled with a mixture of the sample, a ruby chip, and silicone oil as the pressure transmitting medium (PTM). Synchrotron AD-XRD measurements were performed at BL15U1 beamline of Shanghai Synchrotron Radiation Facility (SSRF). The two-dimensional image plate patterns were converted to the one-dimensional intensity versus degree data using the Fit2D software package.[19] The experimental pressures were determined by the pressure-induced fluorescence shift of ruby.[20] The XRD patterns were analyzed with Rietveld refinement using the GSAS program package with a user interface EXPGUI.[21,22]

**Magnetic Properties Measurement**

The magnetization was measured in a Quantum Design magnetic property measurement system (MPMS). Isotherms were collected at an interval of 0.2 K. The magnetic properties may exhibit different behaviors along *ab*-plane and *c*-axis, since the *R*-3 phase of CGT shows a 2D crystalline structure characteristic. Therefore, external magnetic field with the directions parallel and perpendicular to the *c*-axis of the single crystal was applied. Orientations of the crystal were determined on the Laue diffractometer.

**High Pressure Resistance Measurement**

High-pressure resistance measurements were performed using a symmetric DAC with T301 stainless steel gasket. The high-pressure electrical resistance measurement for CGT was performed using a four-electrode method without PTM. To build insulation to the electrodes, the hole in the gasket was filled by the compacted cubic-BN powder and the rest part of the gasket was covered by the insulating gel. The pressure values were determined by monitoring the fluorescence shift of ruby.[20] This configuration for high pressure electrical transport measurement has been successfully applied in our previous work.[17,23]

**The *ab initio* and Molecular Dynamics Simulations**

Our experimental results were supported and further explained by the theoretical simulations. First principles calculations based on density functional theory (DFT), using the projection plane-wave function and pseudopotential method [24,25] with the generalized gradient approximation in the Perdew-Burke-Ernzerhof form (GGA-PBE), were performed with the Vienna Ab initio Simulation Package (VASP) code.[26] The weak van der Waals (vdW) interaction between layers was taken into account by employing DFT-D2 method of Grimme.[27] Due to the localized 3*d* electrons of Cr atoms, the GGA+U (Dudarev) method[28] was employed, where *U-J* = 3.5 eV was referenced to former study.[29] To obtain reasonably stable structure of CGT under different pressures, the lattice parameters and internal atomic positions of the unit cell were fully relaxed until total residual forces were smaller than $10^{-3}$ eV/Å. The reciprocal space sampling was done with Gamma-centered grids of 9×9×3 for the optimization of atomic structure and calculations of electronic structure. The calculated energy bands show a band gap of 0.3 eV under ambient pressure, and the shrinking of band gap upon compression results in a semiconductor-metal transition at around 10 GPa. The theoretical simulation lattice parameter for the ambient phase of CGT is shown in Figure S1, which demonstrates the good agreement between the experimental and theoretical results. This indicted that the selection of parameters in theoretical calculation is appropriate. To mimic the PIA, we performed ab initio molecular dynamics (AIMD) simulations on a large supercell with 270 atoms and relaxed the atoms under high pressure and room temperature. The crystal starts to be amorphized after 8 ps simulation time.

**Results and Disscussion**
**Crystal Structure at Ambient Pressure**

As shown in Figure 1(a), the CGT systems crystallize in a rhombohedral lattice, forming a layered structure stacked along the *c*-axis with a fairly large interlayer spacing of ~ 3.3

Å. Each layer consists of a 2D honeycomb array of Cr atoms in edge-sharing Te octahedrons with the Ge dimers inserted into the resulting octahedral vacancy sites. The Cr moments are normal to the layer, i.e., along the rhombohedral axis. The weak vdW interlayer coupling makes it easy to exfoliate into thin films from bulk crystals.[30] The crystallography phase quality was checked by a Bruker single crystal X-ray Diffractometer. The space group ($R$-3) and lattice parameter ($a = b = 6.8327$ Å, $c = 20.5666$ Å) were obtained after analyzing the single crystal XRD data using the APEX3 package. Our result agrees well with those in earlier reports.[4,5,31] At ambient conditions, CGT crystallizes into a layered crystal structure with space group $R$-3, in which Cr, Ge and Te atoms are located at Wyckoff position 6$c$ (0, 0, 0.3302), 6$c$ (0, 0, 0.0590) and 18$f$ (0.6630, -0.0330, 0.2482), respectively. Each unit cell comprises three layers (formed by CrTe$_6$ and Te$_3$-Ge-Ge-Te$_3$ octahedrons with shared edges) stacked in an $ABC$ sequence along the $c$-axis.[31] Within a filled layer of the octahedron, instead of a random mixture of Cr and Ge, 1/3 of the octahedrons are filled by Ge-Ge dimers (Te$_3$-Ge-Ge-Te$_3$ "dual tripods"), while the other 2/3 are filled by Cr ions. Half of the octahedrons are left empty in Te layers to create a vdW gap.

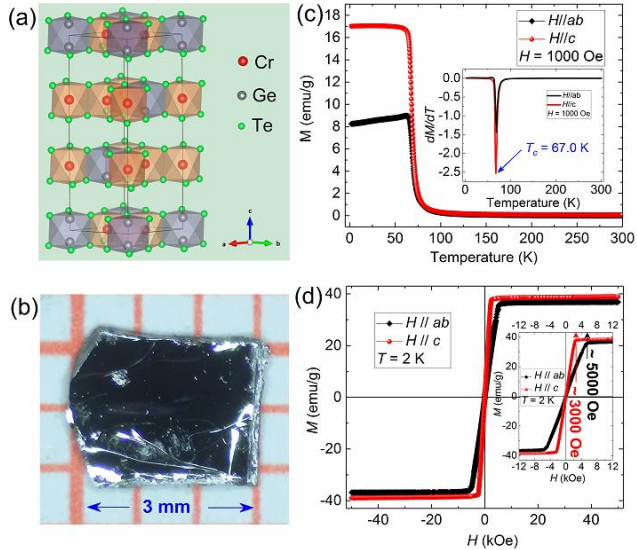

Figure 1. (a) Schematic crystal structure of $R$-3 phase of CGT, which clearly shows the CrTe$_6$ and Te$_3$-Ge-Ge-Te$_3$ octehadrons stacking with edge-sharing along $c$-axis. (b) Optical image of CGT single crystal synthesized in this work. (c) Temperature dependence of magnetization for CGT measured in the magnetic field $H = 1$ kOe. Inset: the derivative magnetization $dM/dT$ vs. $T$, in which the the ferromagnetic transition temperature $T_c \approx 67$ K is roughly determined from the minimum of the $dM/dT$ curve. (d) Field dependence of magnetization for CGT measured at $T = 2$ K. Inset: the magnification of the low-field region, in which the saturation field of $H//ab$ and $H//c$ was determined to be ~ 5000 Oe and ~ 3000 Oe, respectively.

## Magnetic Properties at Ambient Pressure

Figure 1(c) shows the temperature ($T$) dependence of magnetization $M$ ($T$) measured under $H = 1000$ Oe applied parallel to the $ab$ plane and the $c$-axis. A clear paramagnetic (PM) to ferromagnetic (FM) transition is observed. The PM → FM transition temperature $T_c \approx 67$ K is obtained from the minimum of the $dM/dT$ curve as shown in the inset of Figure 1(c), which is in good agreement with the value reported previously.[4,5] Figure 1(d) displays the isothermal magnetization for CGT measured at $T = 2$ K. The enlarged view of the low-field region was shown in the inset of Figure 1(d), in which the saturation fields of $H//ab$ and $H//c$ were determined to be ~ 5000 Oe and ~ 3000 Oe, respectively. This indicates that the $c$-axis is more easily magnetized than that of $ab$-plane, which is consistent with reported results.[4,5] The present magnetic properties measurement results on CGT confirm the good crystalline quality of the crystals used for later measurements.

## Pressure Induced Polymorphic Phase Transition of Cr$_2$Ge$_2$Te$_6$

Figure 2 shows the azimuthally unwrapped XRD images of CGT at different pressures (compression and decompression cycles). The $R$-3 phase (denoted as phase I) retained the ambient pressure crystal structure up to 16.5 GPa. A new diffraction peak appeared at ~ 18.3 GPa (marked with red dashed lines in Figure 3(a)), indicating the occurrence of the pressure-induced crystalline-to-crystalline phase transition. The intensity of the new appeared diffraction peaks for high-pressure phase (denoted as phase II) was enhanced with increasing the pressure. The diffraction peaks meanwhile become rather broadened as the pressure was further increased, indicating the concurrent PIA. With the further increase of pressure up to 35.2 GPa, phase II disappears and the compound completely transforms into the amorphous state. To check whether the AM phase is reversible or not, the XRD measurement during decompression was performed. The diffraction patterns collected at 0 GPa show no sharp peaks, suggesting that the amorphization at high pressure was irreversible.

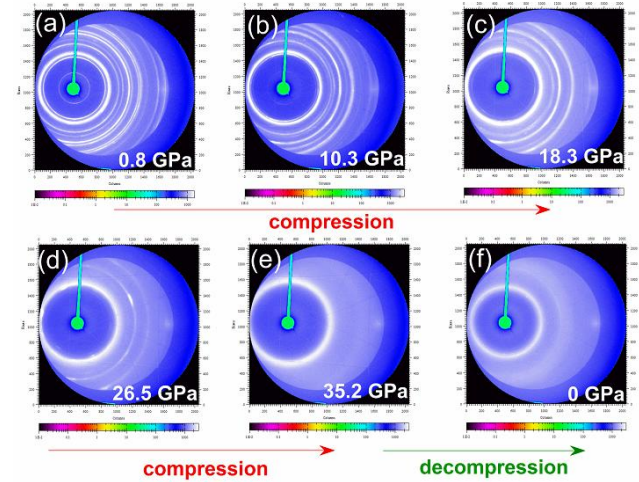

Figure 2. Azimuthally unwrapped XRD images of CGT at different pressures (compression and decompression cycles). A pressure-induced crystallographic phase transition was observed at ~ 18.3 GPa. The XRD peaks become broadened as the pressure increases and eventually the material transforms into the AM state. By releasing the pressure to ambient condition, the AM phase is retained.

The detailed pressure dependence of powder XRD patterns are presented in Figure 3(a). The crystal structure of phase I below 16.5 GPa was refined on the basis of initial model determined by the room temperature single-crystal X-ray diffractometer measurement. Typical GSAS refinement results of CGT under 0.8 and 10.3 GPa are illustrated in the bottom panel of Figure 3(b). Detailed refinement information for phase I is summarized in Table S1. Analysis on the diffraction peaks in Figure 3(a) revealed the coexistence of phase I and II at 18.3 GPa, implying that phase II is probably derived from the atomic rearrangement of phase I. The proportion of phase II increases as the pressure is increased. However, the AM state started to appear (~26.5 GPa) before phase I totally transformed into phase II. Therefore, it is difficult to obtain a pure Phase II under certain pressure. The crystal structure model of Phase II was deduced by testing several candidates. The maximal subgroups of group $R$-3 (148) is $P$-1 (No. 2), $R3$ (146), and $P$-3 (147). At last the crystallographic model of Phase II was analyzed based on drawing on the experience of topological insulator of $Bi_2Te_3$ and crystallographic analysis.[32] The GSAS refinement for phase I and II under 23.4 GPa (top panel of Figure 3(b)) shows good agreement between the experimental data and theoretical models and yields space groups and lattice parameters (phase I: $a$ = 6.4839(13) Å, $c$ = 17.670(5) Å; phase II: $a$ = 3.8091(8) Å, $c$ = 23.260(4) Å). The detailed atomic coordinators for Cr, Ge and Te atoms in phase I and phase II were summarized in Table S2. Phase II is a metastable phase, since it was unquenchable when pressure was released to ambient pressure. This indicated that the formation of the metastable phase under high pressure is a prerequisite for PIA. Phase I is a layered structure with three octahedron layers stacking in the *ABCABC* sequence along the *c*-axis, while Phase II has three octahedron layers stacking in the $A'B'C'A'B'C'$ sequence along the *c*-axis as is shown in Figure 3(c). The $CrTe_6$ and $Te_3$-Ge-Ge-$Te_3$ octehadrons are edge-shared in *ab*-planes for phase I. However, the octahedrons stacking in phase II is more complicated. $CrTe_6$ and $GeTe_6$ octehadrons are edge-shared in *ab*-plane; the $CrTe_6$ and $GeTe_6$ octehadron layers are also edge-shared along the *c*-axis, which is similar to the quintuple layer in topological insulators ($Bi_2Se_3$, $Bi_2Te_3$ and $Sb_2Te_3$)[33] (see Figure 3(c)). Therefore, the stacking sequence for phase II is denser than that of phase I. The polymorphic transition was roughly classified as reconstructive or displacive depending on whether or not breakage of primary interatomic bonding was required in order to interconvert the two crystal phases. The structural transformation (phase I → phase II) in CGT was suggested as a reconstructive-type one.

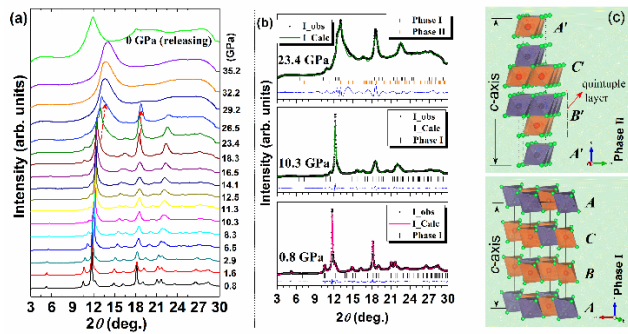

Figure 3. (a) Selected AD-XRD patterns of CGT at room temperature ($\lambda$ = 0.6199 Å). The ambient-pressure $R$-3 structure remains stable up to ~ 18.0 GPa. The appearance of additional diffraction peaks indicates the emergence of second crystalline phase. As the pressure further increases, the diffraction peaks become weak and broadened and eventually the material transforms into the AM state. By decompressing the sample to ambient pressure, the AM phase could be retained. (b) Refined XRD patterns at different pressures ($\lambda$ = 0.6199 Å). The vertical lines denote the theoretical positions of the Bragg peaks. The different curves between observed and calculated XRD patterns are shown at the bottom. (c) Schematic views of crystal structures of phases I and II of CGT, with the red, gray, and green globes represent the Cr, Ge, and Te atoms, respectively.

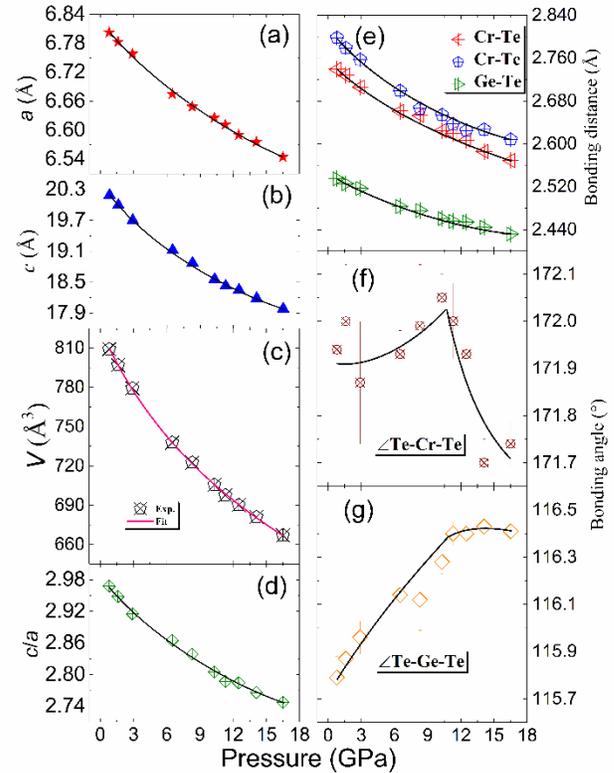

Figure 4. Pressure dependences of lattice parameters (*a* and *c*), axial ratio (*c*/*a*), unit cell volume (*V*), bonding distances (Cr–Te and Ce–Te), and bonding angles (Te–Cr–Te and Te–Ge–Te) for phase I of CGT below 16.5 GPa. The bold lines represent the fitting results to experimental volume by using the third-order Birch-Murnaghan equation of state (BM-EoS).

The pressure dependence of the lattice parameters (*a* and *c*) and lattice volume (*V*) for phase I is displayed in Figures 4(a), (b) and (c). The lattice parameters *a* and *c* homogeneously decrease as the pressure is increased. Figure 4(d) depicts the pressure dependence of axial ratio (*c*/*a*). It is apparent that the *c*-axis is more compressible than that of the *a*-axis, which is related to vdW gaps in phase I along the *c*-axis. Selected bonding distances (Cr-Te and Ge-Te) as a function of pressure are shown in Figure 4(e). Two inequivalent Cr-Te and one Ge-Te bonding distances shrink homogeneously with increasing pres-sure. Instead, the bonding angles ∠Te-Cr-Te (Figure 4(f)) and ∠Te-Ge-Te (Figure 4(g)) show a turning point at ~ 10 GPa. The relative weak vdW bonding is easily compressed with medium pressure. The applied pressure starts to affect the distortion of $CrTe_6$ and (Ge-Ge dimer)-$Te_6$ octahedrons beyond this turning point. The different compressibility of these two octahedrons gives rise to this abnormal bond-angle variation.

**Pressure Induced Amorphization of $Cr_2Ge_2Te_6$**

PIA in solid state materials is not a common phenomenon, in which the crystal lattice is collapsed by applying pressure.[34] Up to now a few materials have been reported to transform into AM state when subjected to high pressure. The most famous case was ice-I (1 GPa and 77 K) reported in 1984.[35] Later the PIA was also observed in other materials such as $SnI_4$ (15-20 GPa) [36,37] and $CH_3NH_3PbBr_3$ (2 GPa).[38] Most experimentally verified PIA was found in oxides such as $Y_2O_3$ (size dependent amorphization, 16 nm size $Y_2O_3$ occurred at 24.8 GPa) [39] and $Ta_2O_5$ (19 GPa).[40] Additionally, chalcogenides such as CuS (18 GPa) [41] and Ge-Sb-Te (GST) [42] were also demonstrated to show the PIA. The mechanism of PIA in oxides was explained as that the applied pressure destroys the polyhedrons of long-range order such as in $Y_2O_3$, while the PIA in GST materials was attributed to the atomic distortion toward disordered vacancies.[7]

One of the most common methods for preparing AM materials is by fast quenching the liquid phase so that the crystallization could be kinetically hindered. In ref 13, the amorphous state of CGT was successfully synthesized by sputtering of Ge, Cr, and Te pure metal targets on a $SiO_2$ (100 nm)/Si substrate. Here, we found that external pressure is an alternative approach to achieve AM phase by creating sufficient lattice distortions in CGT. As stated earlier, the PIA in CGT is irreversible, and hence we provide a new method to prepare the AM CGT other than melt-quench.

AIMD simulation is a computational method for investigating the structural and dynamical properties of materials and has been successfully employed in the study of crystalline-to-crystalline and crystalline-to-AM transitions.[23] The PIA has been observed in crystalline materials with different chemical bonds (hydrogen, vdW, covalent, ionic and metallic bonds) and local crystal structure configurations (compressibility of polyhedrons, voids, etc). Therefore, the structural mechanisms of PIA for specific materials are distinctive. Our AIMD simulations reveal that PIA in CGT is largely induced by random flipping of Ge tripods (similar to the "umbrella flipping" model [43]) toward the vdW gaps, as shown in Figure 5(a). Pressure compresses the Ge-Ge bonds, making it rather rigid. To release the strain energy, one of the Ge atoms has to flip to the other side where the vdW gap offers ample room. The flipping takes place randomly either upward or downward (see the arrows in Figure 5(b)), and Cr/Te atoms are also distorted in a disordered fashion. As a result, the vdW gaps are filled and the long-range order is lost.

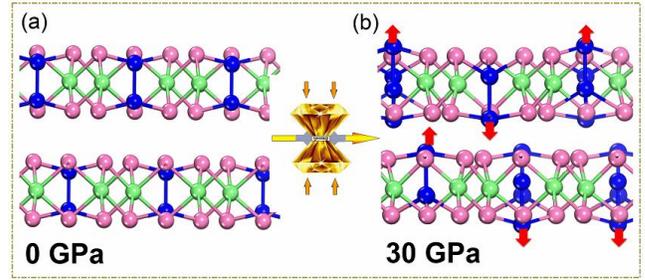

Figure 5. Screenshot of Ge flipping in the Ge-$Te_3$ tetrahedrons by pressure (the left/right panels show the atomic structure (a) before and (b) after the flipping). The arrows indicate the random flipping of Ge atoms toward either side of the vdW gaps.

**Electronic Transport Properties under Extreme Conditions**

It has been demonstrated that the applied pressure plays an important role in influencing the crystal structure and the associated physical properties such as the resistivity contrast in the phase-change memory material $GeSb_2Te_4$.[44] The pressure-induced structural phase transition in CGT is expected to alter its electronic properties, which can be reflected by the measured resistance plotted in Figure 6(a). The resistance shows a rapid decrease for phase I within the measured pressure range. This is because the pressure compresses the vdW gaps first, remarkably reducing the band gap and largely increasing the carrier concentration. The band gap is closed at ~ 10 GPa and the compound undergoes an insulator-to-metal transition (Figure 6(b)). After that, the reduction of resistance slowly goes down until ~ 18 GPa, at which a bouncing point is observed. This abnormal point in resistance is apparently correlated to the emergence of new phases in CGT. The atomic disorder in AM phase could enhance the phonon-electron scattering and the localization of electron wave function [45] resulting in a slight increase in resistance.

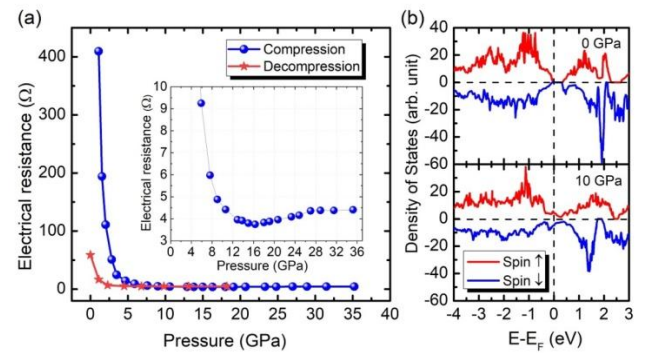

Figure 6. (a) Resistance of sample during compression and decompression. The resistance of CGT precipitously decreases with pressure by ~2 orders of magnitude below 10 GPa due to the shrinking band gap (b). A bouncing point was observed around ~ 18 GPa as shown in the zoom-in inset, corresponding to the phase transition (phase I→phase II/AM). Upon decompression, the amorphous phase is retained and its resistance increases. After unloading the sample to ambient pressure, the resistance is abnormally smaller than that of the initial crystal.

Upon decompression, the resistance increases subtly as the pressure is decreased to ~ 2 GPa. Matter in the AM state does not own long-range order and thus was expected to show larger resistance compared with crystalline counterpart (under ambient conditions, the AM phase of PCMs normally exhibits higher resistance by 2 – 5 orders of magnitude than the crystal). However, the resistance of CGT after decompressed to ambient conditions in this work fails to exceed the initial crystalline phase. This indicates that the CGT has a unique AM phase different from other PCMs such as GST.

**Conclusions**

To summarize, we performed *in situ* high-pressure AD-XRD experiments on CGT crystals. A mixture of polymorphic transition and amorphization is observed at ~18 GPa. At higher pressure around 29 GPa, all the crystalline phases transform into the AM one. The PIA in this material is achieved via the stochastic flipping of Ge in the Ge-Te$_3$ tripods toward vdW gaps, different from the reported PCMs which is realized by random vacancy occupation. In addition, the electronic property was investigated under high-pressure and room temperature and a bouncing point was observed in the resistance change when new phases appear. The AM phase becomes metallic upon further compression and the resistance shows subtle pressure dependence. By unloading the sample to the ambient pressure, the AM state shows an abnormally lower resistance than the initial crystal. This study provides an experimental approach to tune the electronic structure of 2D-like magnetic materials by applying strain on them and contributes a new way to synthesize the AM phase.

ACKNOWLEDGEMENTS

ASSOCIATED CONTENT

**Supporting Information.**
The Supporting Information is available free of charge on the ACS Publications website DOI: 10.1021/ACS.JPCC.9B02415.

Atomic coordination parameters of phase I of Cr$_2$Ge$_2$Te$_6$ below 16.4 GPa (Table S1), atomic coordination parameters of Cr$_2$Ge$_2$Te$_6$ at 23.4 GPa (Table S2), theoretical simulation lattice parameter results of Phase I Cr$_2$Ge$_2$Te$_6$ (Figure S1), and the complete author list for Refs 2, 3, 4, 6, 10, 16, 17, 29, 39.

AUTHOR INFORMATION

Corresponding Author
*Authors to whom correspondence should be addressed.
E-mail address: mxu@hust.edu.cn (MX), zhaojinggeng@163.com (JGZ), and guoyf@shanghaitech.edu.cn (YFG).

Author Contributions
#Z.H.Y, and W.X. contributed equally.

Acknowledgements.

The authors acknowledge the support by the National Science Foundation of China (No.11874264), the Natural Science Foundation of Shanghai (No. 17ZR1443300), the Shanghai Pujiang Program (No. 17PJ1406200), and the Natural Science Foundation of Heilongjiang Province (Grant No. A2017004). M.X. and X.M. acknowledge the National Key R&D Plan of China (2017YFB0701701 and 2017YFB0405601). High-pressure synchrotron XRD work was performed at the BL15U1 beamline, Shanghai Synchrotron Radiation Facility (SSRF) in China. We thank Lili Zhang, Chunyin Zhou, Shuai Yan and Ke Yang (BL15U1) for assistance with high-pressure XRD measurements. The authors declare no competing financial interest.

# *Supporting information*

# Pressure-induced Structural Phase Transition and a Special Amorphization Phase of Two-dimensional Ferromagnetic Semiconductor $Cr_2Ge_2Te_6$

Zhenhai Yu[†,#], Wei Xia[†,‡,#], Kailang Xu[§,#], Ming Xu[§,*], Hongyuan Wang[†,‡], Xia Wang[∥], Na Yu[∥], Zhiqiang Zou[∥], Jinggeng Zhao[⊥,*], Lin Wang[⊥], Xiangshui Miao[§], and Yanfeng Guo[†,‡,*]

[†]School of Physical Science and Technology, ShanghaiTech University, Shanghai 201210, China

[‡]University of Chinese Academy of Sciences, Beijing 100049, China

[§] School of Optical and Electronic Information, Huazhong University of Science and Technology, Wuhan 430074, China

[∥] Analytical Instrumentation Center, School of Physical Science and Technology, ShanghaiTech University, Shanghai 201210, China

[⊥]Department of Physics, Harbin Institute of Technology, Harbin 150080, China

[⊥]Center for High Pressure Science and Technology Advanced Research, Shanghai, 201203, China

#These authors contributed equally to this study and share first authorship.

*Corresponding author: E-mail address: mxu@hust.edu.cn (M. Xu),

zhaojinggeng@163.com (J. G. Zhao), and guoyf@shanghaitech.edu.cn (Y. F. Guo).

**Table S1**. Atomic coordination parameters of phase I of $Cr_2Ge_2Te_6$ below 16.4 GPa (Space group: $R$-3; Cr: 6$c$; Ge: 6$c$; Te: 18$f$)

| $P$ (GPa) | $a$ (Å) | $c$ (Å) | $z_{Cr}$ | $z_{Ge}$ | $x_{Te}$ | $y_{Te}$ | $z_{Te}$ | $R_{wp}$ | $R_p$ | $\chi^2$ |
|---|---|---|---|---|---|---|---|---|---|---|
| 0.8 | 6.8027(2) | 20.189(3) | 0.3299(6) | 0.0590(3) | 0.6633(3) | 0.9672(2) | 0.24821(1) | 4.59% | 3.45% | 1.253 |
| 1.6 | 6.7830(4) | 19.998(3) | 0.3302(5) | 0.0591(2) | 0.6633(2) | 0.9673(3) | 0.24821(1) | 7.73% | 4.96% | 1.689 |
| 2.9 | 6.7583(5) | 19.701(4) | 0.3301(6) | 0.0591(2) | 0.6633(2) | 0.9670(3) | 0.24821(1) | 9.51% | 6.54% | 2.983 |
| 6.5 | 6.6749(4) | 19.123(2) | 0.3304(2) | 0.0593(1) | 0.6632(2) | 0.9671(2) | 0.24821(1) | 8.78% | 5.98% | 2.876 |
| 8.3 | 6.6486(6) | 18.874(4) | 0.3316(6) | 0.0589(5) | 0.6634(4) | 0.9667(4) | 0.24821(1) | 9.81% | 6.87% | 4.669 |
| 10.3 | 6.6258(5) | 18.558(3) | 0.3308(2) | 0.0593(2) | 0.6635(1) | 0.9676(1) | 0.24817(1) | 7.58% | 5.40% | 3.280 |
| 11.3 | 6.6115(6) | 18.428(5) | 0.3312(4) | 0.0595(2) | 0.6631(2) | 0.9671(2) | 0.24821(1) | 6.49% | 4.78% | 3.654 |
| 12.5 | 6.5901(6) | 18.350(4) | 0.3330(4) | 0.0594(2) | 0.6637(2) | 0.9664(2) | 0.24822(2) | 5.57% | 4.22% | 3.066 |
| 14.1 | 6.5756(5) | 18.187(3) | 0.3301(2) | 0.0595(2) | 0.6632(2) | 0.9667(2) | 0.24818(3) | 4.94% | 3.74% | 2.904 |

| 16.5 | 6.5447(7) | 17.983(4) | 0.3302(2) | 0.0592(1) | 0.6632(1) | 0.9669(2) | 0.24820(2) | 5.60% | 4.15% | 2.944 |

$$R_p = \sum(y_{oi} - y_{ci}(\alpha)) / \sum y_{oi}$$

$$R_{wp} = \sqrt{\sum w_i(y_{oi} - y_{ci}(\alpha))^2 / \sum w_i y_{oi}^2}$$

$$\chi^2 = (1/N)\sum_i (y_{oi} - y_{ci})^2 / \sigma^2[y_{oi}]$$

$y_{oi}$ is the experimental observed intensity of XRD, $y_{ci}$ is the calculated intensity of XRD, $\alpha$ is the variable in refinement, $W_i$ is weight factor.

**Table S2**. Atomic coordination parameters of $Cr_2Ge_2Te_6$ at 23.4 GPa ($R_p$ = 3.13%, $R_{wp}$ = 4.84%, $\chi^2$ = 1.323)

| Atom | Site | x | y | z |
|---|---|---|---|---|
| **Phase I** | | | | |
| Space group: *R*-3; *a* = 6.4839(13) Å, *c* = 17.670(5) Å. | | | | |
| Cr | 6*c* | 0 | 0 | 0.3312(2) |
| Ge | 6*c* | 0 | 0 | 0.0602(2) |
| Te | 18*f* | 0.6625(1) | 0.9642(1) | 0.2483(1) |
| **Phase II** | | | | |
| Space group: *R*3; *a* = 3.874(2) Å, *c* = 23.631(12) Å. | | | | |
| Cr | 3*a* | 0 | 0 | 0.605(3) |
| Ge | 3*a* | 0 | 0 | 0.393(3) |
| Te(1) | 3*a* | 0 | 0 | 0.797(2) |
| Te(2) | 3*a* | 0 | 0 | 0.196(2) |
| Te(3) | 3*a* | 0 | 0 | 0 |

**The theoretical simulation lattice parameter results of Phase I Cr$_2$Ge$_2$Te$_6$**

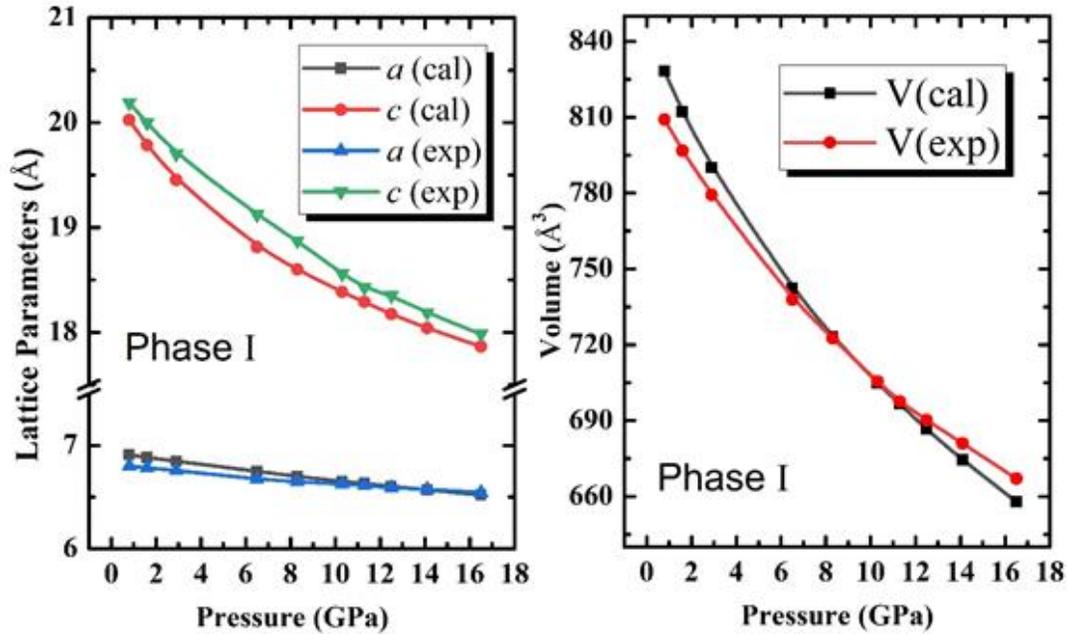

**Figure S1.** The theoretical simulation lattice parameter for Phase I Cr$_2$Ge$_2$Te$_6$, in which shows the good agreement between the experimental and theoretical results. This indicted that the selection of parameters in theoretical calculation is appropriate

**The complete author list:**

TOC graphic

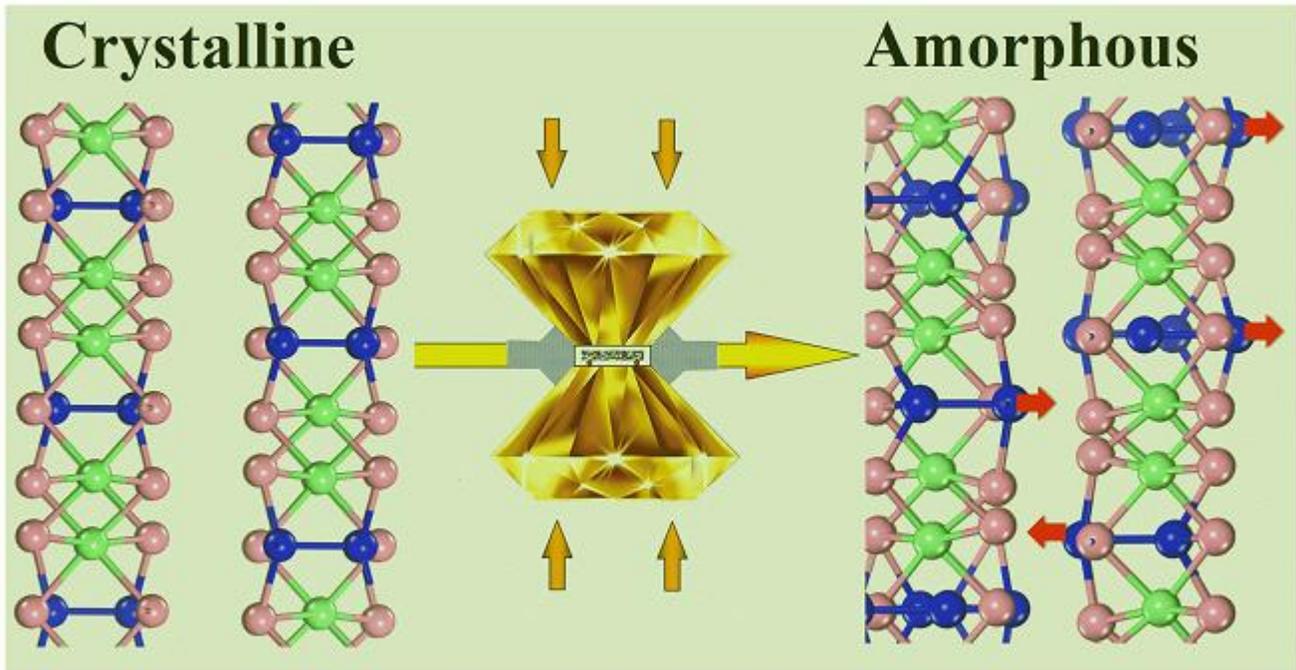